\documentclass[a4paper,twocolumn]{article}
\usepackage[utf8]{inputenc}
\usepackage[T1]{fontenc}
\usepackage[numbers,compress,merge]{natbib}
\usepackage[colorlinks=true]{hyperref}
\usepackage{graphicx}
\usepackage{amsmath,amssymb}
\usepackage{subfigure}
\usepackage{color,xcolor}
\usepackage[varg]{txfonts}
\usepackage[affil-it]{authblk}
\title{Tidal-charge effects on the superradiance of rotating black holes}

\author{Ednilton S. de Oliveira\footnote{ednilton@ufpa.br}}
\affil{Faculdade de F\'isica, Universidade Federal do Par\'a,
	66075-110, Bel\'em, Par\'a, Brazil}

\date{\today}

\begin{document}

\maketitle

\begin{abstract}
The changes a (negative) tidal charge causes at the phenomenon of superradiance which occurs around rotating black holes are investigated. This is made by computing the amplification factors of massless scalar waves being scattered by the black hole. It is shown that the increase of the tidal charge intensity leads to a considerable enhancement of energy extraction from near-extreme black holes. Such improvement results from the fact that extreme black holes with more negative tidal charges spin faster. Maximum amplification decreases with the increase of the tidal charge intensity if the angular momentum of the black hole per unit mass is fixed. The tidal charge may also change crucially the superradiance phenomenon of massless scalar waves causing maximum amplification to occur for $m > 1$ differently from the case of Kerr black holes.
\end{abstract}

\section{Introduction}

A couple of decades ago it has been proposed the existence of black holes with tidal charge~\cite{Dadhich2000plb487_1} in the context of the Randall-Sundrum brane-world scenario~\cite{Randall1999prl83_3370,Randall1999prl83_4690}. This charge results from the influence of the fifth dimension in our 4-dimensional observed universe, the brane. The exclusive effect of the tidal charge on the black hole physics has been quantified in several researches based on the study of static black holes~\cite{Molina2016prd93_124068,Toshmatov2016prd93_124017,Abdujabbarov2017prd96_084017,Rahimov2011ass335_499,deOliveira2018epjc78_876,deOliveira2019}, among which our works  treating about wave scattering from such black holes~\cite{deOliveira2018epjc78_876,deOliveira2019}. In these papers we have shown that black holes endowed with more intense (negative) tidal charges\footnote{We consider only black holes with negative tidal charges since these, as argued in Ref.~\cite{Dadhich2000plb487_1}, are physically more natural than the ones with positive tidal charges.} present bigger absorption cross sections, while their scattering spectra in the weak-field regime is not easily distinguishable from that of Schwarzschild black holes. This leads to the conclusion that the tidal charge
interferes considerably in the dynamics near the black hole, but with an influence which can be neglected in the far region.


When the effect of the tidal charge is combined with other properties, as angular momentum and electric charge~\cite{Aliev2005prd71_104027}, the black hole presents novel configuration compared with black holes predicted within General Relativity. Specially in the case of electrically neutral rotating black holes, the tidal charge modifies the limit of angular momentum such black holes can acquire, allowing them to spin faster than extreme Kerr black holes. Also, some of these black holes possess a bigger ergoregion, what could allow more energy extraction from them than from Kerr black holes. Indeed, it has been recently shown that a negative tidal charge increases the efficiency of energy extraction from the black hole via the Penrose process~\cite{Khan2019pdu26_100331}. Still concerning about rotating black holes, the tidal charge has also important implications to astrophysics, as it may alters the radii size of ISCO~\cite{Aliev2005prd71_104027,Abdujabbarov2010prd81_044022,Pun2008prd78_084015} and of the shadow cast by the black hole~\cite{Abdujabbarov2017prd96_084017,Schee2008ijmpd18_983,Amarilla2011prd85_064019,Eiroa2017epjc78_91}.

The existence of an ergoregion allows energy extraction from a rotating black hole via the Penrose process~\cite{Penrose1969nc1_252} or superradiance (see Ref.~\cite{Brito2015lnp906} for a review). The latter occurs when a bosonic wave is scattered by the black hole. If the wave has a specific set of properties, then it can emerge from such scattering with more energy than it came in. It's worth mentioning that rotational superradiance is not an exclusivity of black holes, but the first experimental observation in laboratory of this phenomenon has been recently preformed in a draining vortex~\cite{Torres2017np13_833}, an analog~\cite{Unruh1981prl46_1351,Visser1997cqg15_1767} of rotating black holes.

The increase on the intensity of a (negative) tidal charge makes the ergoregion expand. This indicates that the presence of the tidal charge on rotating black holes alters quantitatively the process of superradiance. However, the increase of the ergoregion is not enough to lead us to conclude that maximum amplification of a scattered wave will get enhanced in the same sense. The size of the event horizon plays also an important role in the quantity of energy which can be extracted from the black hole. Once the tidal charge also changes the event horizon size, the increase of the tidal-charge intensity may result in an increase or decrease of maximum amplification obtained in the scattering process.

In this work we compute the amplification factors of massless scalar waves which are scattered by rotating black holes with tidal charge. We show that the tidal charge acts to enhance considerably the efficiency of the superradiance of near-extreme black holes. On the other hand, maximum amplification gets suppressed by the increase of the tidal-charge intensity if we consider black holes of fixed angular momentum/mass ratio. We present and compare both low-frequency approximation and numeric results for the amplification factors showing they are in excellent agreement.

This paper is divided as follows: in Sec.~\ref{sec:class_desc} we briefly review the main properties of rotating black holes with tidal charge; in Sec.~\ref{sec:wave_eq} we present the equations of the massless scalar waves around the considered black holes as well as an analytic approach which allows us to obtain amplification factors at low frequencies; Section~\ref{sec:results} provides a selection of results obtained numerically, and also their comparisons with the low-frequency approximation; our conclusions are presented in Sec.~\ref{sec:conclusions}. We work with units $c = G = \hbar = 1$.

\section{Black hole description}
\label{sec:class_desc}

The line element of rotating black holes with tidal charge in Boyer-Lindquist coordinates is~\cite{Aliev2005prd71_104027}:

\begin{eqnarray}
ds^2 & = & \left(1-\frac{2Mr-\beta}{\rho^2}\right) dt^2 + 2a\frac{2Mr-\beta}{\rho^2} \sin^2\theta \, dt d\phi \nonumber \\ &&  - \frac{\rho^2}{\Delta} dr^2 - \rho^2 d\theta^2\nonumber \\
&& - \left(r^2 + a^2 + \frac{2Mr-\beta}{\rho^2} a^2 \sin^2\theta \right) \sin^2 \theta\, d\phi^2,
\label{ds}
\end{eqnarray}
where
$$ \Delta = r^2 -2Mr + a^2 + \beta,$$
$$ \rho^2 = r^2 + a^2\cos^2\theta,$$
$M$ is the mass of the black hole, and $a$ is the ratio between its angular momentum and mass. The parameter $\beta \equiv qM^2$ may be negative in the context of the Randall-Sundrum brane-world scenario. However, if $ 0 < \beta \le M^2-a^2 $, the metric above represents the Kerr-Newman solution~\cite{Newman1965jmp6_918}, which describes the spacetime around electrically charged rotating black holes. The roots of $\Delta$ are:
\begin{equation}
 r_{\pm} = M\pm\sqrt{M^2-a^2 - \beta},
 \label{horizons}
\end{equation}
with $r_+$ determining the event horizon position. From this we can see that naked singularities occur if $ a^2 > M^2 - \beta$, and extreme black holes have $a_c = \sqrt{M^2 - \beta}$. For negative values of $\beta$, extreme black holes have higher angular momentum as higher is the value of $|\beta|$.

The ergoregion of rotating black holes with tidal charge is limited by the surface located at:
\begin{equation}
 r_\text{ergo} (\theta)= M + \sqrt{M^2-a^2\cos^2\theta -\beta}.
 \label{ergo}
\end{equation}
Since $\beta < 0$, rotating black holes with tidal charge have an ergoregion bigger then Kerr black holes with same angular momentum. A bigger ergoregion is not enough to determine if more energy can be extracted from the black hole. The key feature for superradiance of rotating black holes is the space between the ergosurface and the event horizon.

\begin{figure}
 \centering
 \includegraphics[width=0.49\textwidth]{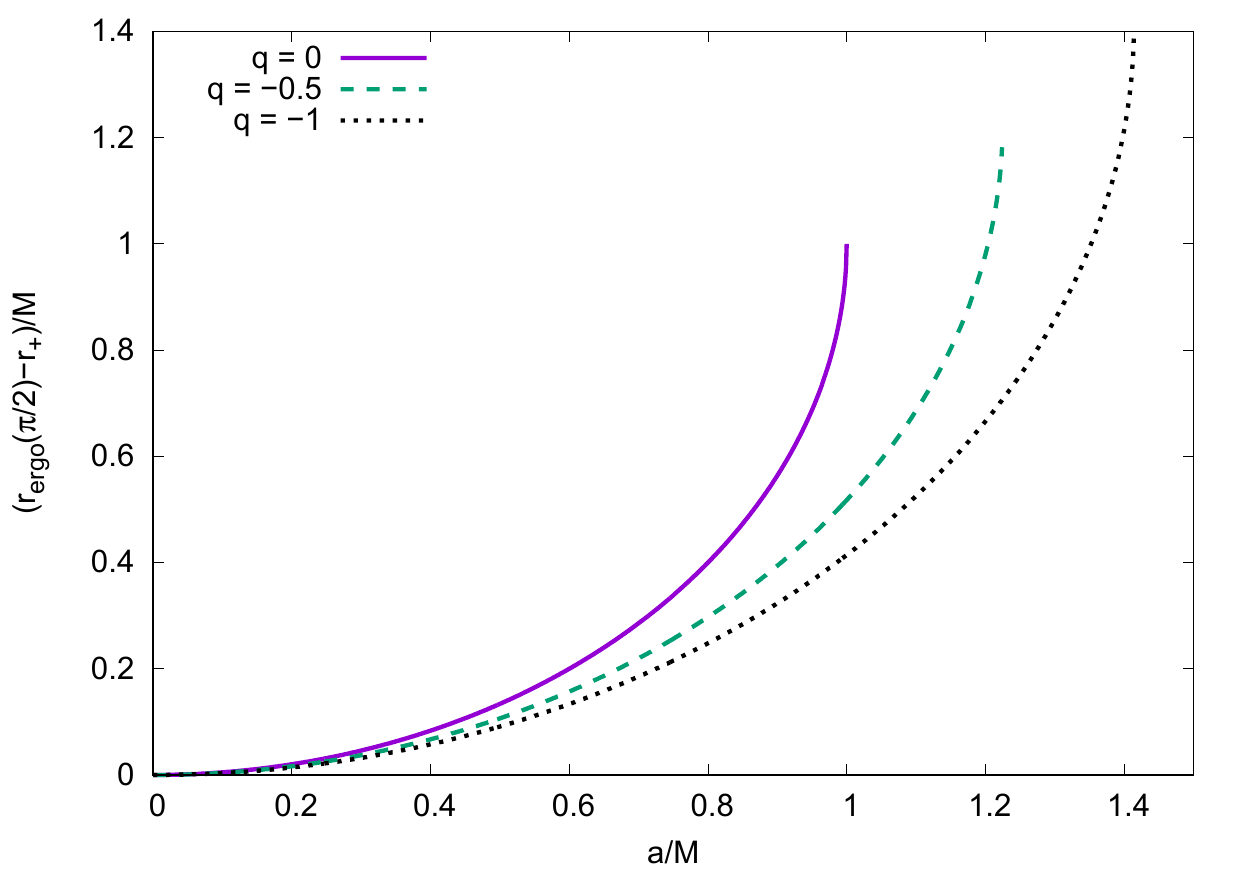}
 \caption{Size difference between the ergosurface and the event horizon at the black hole equator. This illustrates that although ergoregion increases with the tidal-charge intensity, the space between the ergosurface and the event horizon of black holes with tidal charge is smaller than that of Kerr black holes with same angular momentum.}
 \label{fig:erg_rh}
\end{figure}

In Fig.~\ref{fig:erg_rh} we show how the size difference between the ergoregion and the event horizon at the black hole equator changes with the angular momentum for black holes with different tidal charges. Although the increase of the tidal-charge intensity induces an increase of the ergoregion, the event horizon increases faster in the same sense, resulting in a reduction of space between the event horizon and the ergosurface for black holes with same $a$. On the other hand, if $a$ is allowed to increase, then the space between the event horizon and the ergosurface may be bigger in the cases of near-extreme black holes with tidal charge than in the case of extreme Kerr black holes. To make this clearer, let us express the size difference between the ergosurface and the event horizon at the equator of extreme black holes. This is:
\begin{equation}
r_\text{ergo}(\pi/2) -r_+ = \sqrt{M^2-\beta},
\label{erg_rh_eq}
\end{equation}
which evidently increases with the increase of $-\beta$.

Another important property of rotating black holes is the angular velocity at their event horizon:
\begin{equation}
\Omega_H \equiv \frac{a}{r_+^2 + a^2}.
\label{Omega}
\end{equation}
This angular velocity plays an important role in the phenomenon of superradiance as its value determines, together with the azimuthal quantum number, the maximum frequency a wave must have to get amplified in the scattering. We note that the angular velocity at the event horizon of extreme black holes is $\Omega_{Hc} = \sqrt{M^2-\beta}/(2M^2-\beta )$, meaning that extreme rotating black holes with tidal charge have smaller angular velocities, although they have higher angular momentum than Kerr black holes of same mass.

\section{Massless scalar field}
\label{sec:wave_eq}

\subsection{Wave equations}

Here we use the model of the massless scalar field to study the effect of the tidal charge on superradiance. This field obeys the Klein-Gordon equation given by:
\begin{equation}
 \frac{1}{\sqrt{-g}} \partial_{\mu} \left ( \sqrt{-g} g^{\mu\nu} \partial_\nu \Phi \right)  = 0.
 \label{KG}
\end{equation}

A separation of variables is possible for this equation if we assume:
$$
\Phi = \frac{R_{\omega lm}(r)}{\sqrt{r^2+a^2}} S_{\omega lm}(\theta) e^{i(m\phi-\omega t)}.
$$
The functions $S_{\omega lm}$ are the oblate spheroidal harmonics~\cite{Abramowitz_etal1964}, which obey the equation:
\begin{equation}
\frac{d}{dx} \left((1-x^2) \frac{dS_{\omega lm}}{dx} \right) + \left[(a\omega)^2 x^2 - \frac{m^2}
{1-x^2} + \lambda_{lm}\right] S_{\omega lm} = 0,
\label{S}
\end{equation}
where $x = \cos\theta$, and $\lambda_{lm}$ are the eigenvalues of the $S_{\omega lm}$.

The radial equation resulting from the variable separation can be put in the form:
\begin{equation}
\frac{d^2}{dr_*^2}R_{\omega lm} + \left[\omega^2 - V_{\omega lm}(r)\right]R_{\omega lm} = 0,
\label{radial_eq}
\end{equation}
where $r_*$ is the tortoise coordinate, defined as
$$
\frac{d}{dr_*} = \frac{\Delta}{r^2+a^2} \frac{d}{dr},$$
and the effective potential is
\begin{eqnarray}
V_{\omega lm} (r) & = & -\frac{1}{(r^2+a^2)^2} \left[m^2a^2-\Delta(\lambda_{lm}+\omega^2a^2) \right. \nonumber\\
&& \left. + 2ma\omega(\beta - 2Mr)\right] + \Delta\frac{\Delta+2r(r-M)}{(r^2+a^2)^3} \nonumber\\
&& -\frac{3r^2\Delta^2}{(r^2+a^2)^4}.
\label{V}
\end{eqnarray}

Analytic solutions to the radial equation can be easily found in the near-horizon region and far from the black hole. Considering we are dealing with a scattering problem of a wave incoming from infinity, for $r \gtrsim r_+$ we have:
\begin{equation}
R_{\omega lm} \sim \mathcal{T} e^{-i \tilde{\omega} r_*},
\label{asymp_rh}
\end{equation}
where $\tilde{\omega} = \omega - m\Omega_H$ and $\mathcal{T}$ is the transmitted amplitude at the event horizon. For $r \to \infty$:
\begin{equation}
R_{\omega lm} \sim  \mathcal{I} e^{-i \omega r_*} +  \mathcal{R} e^{i \omega r_*},
\label{asymp_inf}
\end{equation}
where $\mathcal{I}$ and $\mathcal{R}$ are the incident and reflected amplitudes at infinity. The reflection and transmission rates are given respectively by $|\mathcal{R}|^2/|\mathcal{I}|^2$ and $|\mathcal{T}|^2/|\mathcal{I}|^2$. These rates are related to each other by:
\begin{equation}
 \frac{|\mathcal{R}|^2}{|\mathcal{I}|^2} = 1 - \frac{\tilde{\omega}}{\omega} \frac{|\mathcal{T}|^2}{|\mathcal{I}|^2}.
 \label{coefs_rel}
\end{equation}

The greybody factors, which mesure the ratio between absorbed and incident fluxes, are:
\begin{equation}
\gamma_{lm} = 1- \frac{|\mathcal{R}|^2}{|\mathcal{I}|^2}.
\label{gbf}
\end{equation}
The fact that $\tilde{\omega} < 0$ when $\omega < m\Omega_H$ leads the reflection coefficient to be higher than unity in such regime, i. e., the wave gets amplified in the scattering process. In this regime $\gamma_{lm} < 0$, and we describe the phenomenon of superradiance in terms of the amplification factor:
\begin{equation}
Z_{lm} = \frac{|\mathcal{R}|^2}{|\mathcal{I}|^2} - 1.
\label{Z}
\end{equation}

The reflection coefficient is obtained by using the asymptotic matching technique, which relies on comparing the solution of the radial equation~\eqref{radial_eq} with its asymptotic approximations~\eqref{asymp_rh} and~\eqref{asymp_inf}. Here we evaluate the radial equation both in the low-frequency regime and numerically to determine the reflection coefficients. Although asymptotic form~\eqref{asymp_inf} is useful to clearly express the scattering process, its improvement usually helps to reach numeric convergence when applying the matching technique. One way of improving~\eqref{asymp_inf} is, as done for example in Ref.~\cite{Macedo2013prd88_064033}, writing:
$$
R_{\omega lm} \sim \mathcal{I} e^{-i\omega r_*} \sum\limits_{j=0}^N \frac{A_j}{r^j} + \mathcal{R} e^{i\omega r_*} \sum\limits_{j=0}^N \frac{A^*_j}{r^j},
\label{asymp_series}
$$
where the coefficients $A_j$ are obtained by reinserting the above expression into the radial equation and $N$ is chosen according to the desired precision.

An alternative can be obtained by noting that for $r \gg r_+$ ($r_* \sim r$), the radial equation can be approximated to:
\begin{equation}
 \frac{d^2}{dr_*^2}R_{\omega lm} + \left(\omega^2 - \frac{a^2\omega^2+\lambda_{lm}}{r_*^2}\right)R_{\omega lm} = 0,
 \label{rad_eq_asy}
\end{equation}
which solution is
\begin{equation}
R_{\omega l m} \approx \sqrt{\frac{\pi\omega r_*}{2}}\left[(-i)^{\nu+1/2} \mathcal{I} H_{\nu}^{(2)} (\omega r_*) + i^{\nu+1/2} \mathcal{R} H_\nu^{(1)} (\omega r_*) \right],
\label{asy_hankel}
\end{equation}
where $H_\nu^{(1),(2)}$ are the Hankel functions~\cite{Abramowitz_etal1964}, and $\nu = \sqrt{\lambda_{lm} + a^2\omega^2 + 1/4}$. Using the asymptotic expansions $H^{(1)}_\nu(x) \sim \sqrt{2/(\pi x)} e^{i(x-\nu\pi/2-\pi/4)}$ and $H^{(2)}_\nu(x) \sim \sqrt{2/(\pi x)} e^{-i(x-\nu\pi/2-\pi/4)}$, it is straightforward to show that~\eqref{asy_hankel} reduces to~\eqref{asymp_inf} for $\omega r_* \to \infty$. Here we use asymptotic form~\eqref{asy_hankel} to compute the amplification factors both in the low-frequency regime and numerically.

\subsection{Low-frequency approximation}

A low-frequency approximation can be found by making $\omega = 0$ in the radial equation, and then matching its solution with the asymptotic form~\eqref{asy_hankel} for $\omega r_* \ll 1$. Considering that $\lambda_{lm} \approx l(l+1) +\mathcal{O}(\omega^2a^2)$ in the regime $\omega a \ll 1$~\cite{Abramowitz_etal1964}, and that we are dealing with a scattering problem, the low-frequency solution for Eq.~\eqref{radial_eq} we seek is ($ma \neq 0$):
\begin{equation}
 R_{\omega lm} \approx \frac{C_1}{r_+-r_-} \left(\frac{r-r_+}{r-r_-}\right)^{i\eta} \sqrt{r^2+a^2} \,F (-l,l+1;2i\eta+1;-z),
 \label{zero_freq_sol}
\end{equation}
where $F(\alpha,\beta;\gamma;x)$ are hypergeometric functions~\cite{Abramowitz_etal1964}, $\eta \equiv ma/(r_+-r_-)$, and
$$ z \equiv \frac{r-r_+}{r_+-r_-}.$$

The expansion of~\eqref{zero_freq_sol} in the regime $r\gg r_+$ is:
\begin{eqnarray}
 R_{\omega l m} & \sim & C_1 \Gamma(2i\eta+1) \left[\frac{(2l)!}{l!\Gamma(2i\eta+l+1)}\left(\frac{r}{r_+-r_-}\right)^{l+1} \right. \nonumber\\
 && \left. + \frac{l!}{2(-1)^{l+1}(2l+1)!\Gamma(2i\eta-l)}\left(\frac{r}{r_+-r_-}\right)^{-l}\right].
 \label{zero_freq_asy}
\end{eqnarray}

On the other hand, using that $H_\nu^{(1)}(x) = J_\nu(x)+iN_\nu(x)$, and $H_\nu^{(2)} = J_\nu(x)-iN_\nu(x)$, where $J_\nu(x)$ and $N_\nu(x)$ are the Bessel and Neumann functions, respectively, we can obtain that in the regime $\omega r \ll 1$~\eqref{asy_hankel} reduces to:
\begin{eqnarray}
R_{\omega l m} & \sim & [(-i)^{l+1}\mathcal{I} + i^{l+1} \mathcal{R}] \frac{\sqrt{\pi}(\omega r)^{l+1}}{2^{l+1}(l+1/2)!} \nonumber\\ 
&& + i[(-i)^{l+1}\mathcal{I} -i^{l+1}\mathcal{R}]\frac{2^l(l-1/2)!}{\sqrt{\pi}(\omega r)^l}.
\label{asy_lf}
\end{eqnarray}

Both expressions~\eqref{zero_freq_asy} and~\eqref{asy_lf} describe the wave in the low-frequency regime in the far region. Matching the corresponding terms, after some manipulation, we can arrive at:
\begin{equation}
 \left|\frac{\mathcal{R}}{\mathcal{I}}\right|^2 \approx \left|\frac{1+X}{1-X}\right|^2,
 \label{lf_R}
\end{equation}
where
\begin{equation}
 X = i\frac{(-1)^{l+1}2^{2l-1}}{(2l+1)^2}\left[\frac{l!}{(2l)!}\right]^4\frac{\Gamma(2i\eta+l+1)}{\Gamma(2i\eta-l)}[\omega(r_+-r_-)]^{2l+1}.
\end{equation}

Approximation above is valid only for $ma \neq 0$. For $m = 0$, the solution for the radial equation with $\omega = 0$ is different of~\eqref{zero_freq_sol}. However, since we focus on superradiance, which occurs only for $m > 0$, an analysis in the low-frequency regime for $m = 0$ reveals irrelevant here. Nevertheless, this low-frequency approximation is also valid for non-superradiant scattering with $m < 0$, so that we also compare it with our results for the greybody factors. As we see in the next section, this approximation excellently agrees with the numeric results for $M\omega \lesssim 0.01$.

\section{Numeric results}
\label{sec:results}

Since the analytic results for the reflection coefficients presented in the previous section are only valid for small frequencies, our main results are computed numerically. First we find the oblate spheroidal harmonics eigenvalues $\lambda_{lm}$ numerically~\cite{Hughes2000prd61_084004,Dolan2008cqg25_235002}, and then insert them in the radial equation~\eqref{radial_eq}. The radial equation is evaluated via fourth-fifth order Runge–Kutta method with boundary conditions given by the near-horizon solution~\eqref{asymp_rh} at $r = (1+\epsilon) r_+$ ($\epsilon \ll 1$). The resulting solution is matched with the far-region solution~\eqref{asy_hankel} in terms of Hankel functions for $r \gg r_+$ in order to find the appropriate coefficients. Precision of our results has been verified by comparing them with the low-frequency approximation~\eqref{lf_R}, and by monitoring if flux conservation~\eqref{coefs_rel} was maintained apart from a small tolerated error upon numerical evaluation of the radial equation.

\subsection{Superradiant regime}

\begin{figure}[!htb]
\centering
\includegraphics[width=0.49\textwidth]{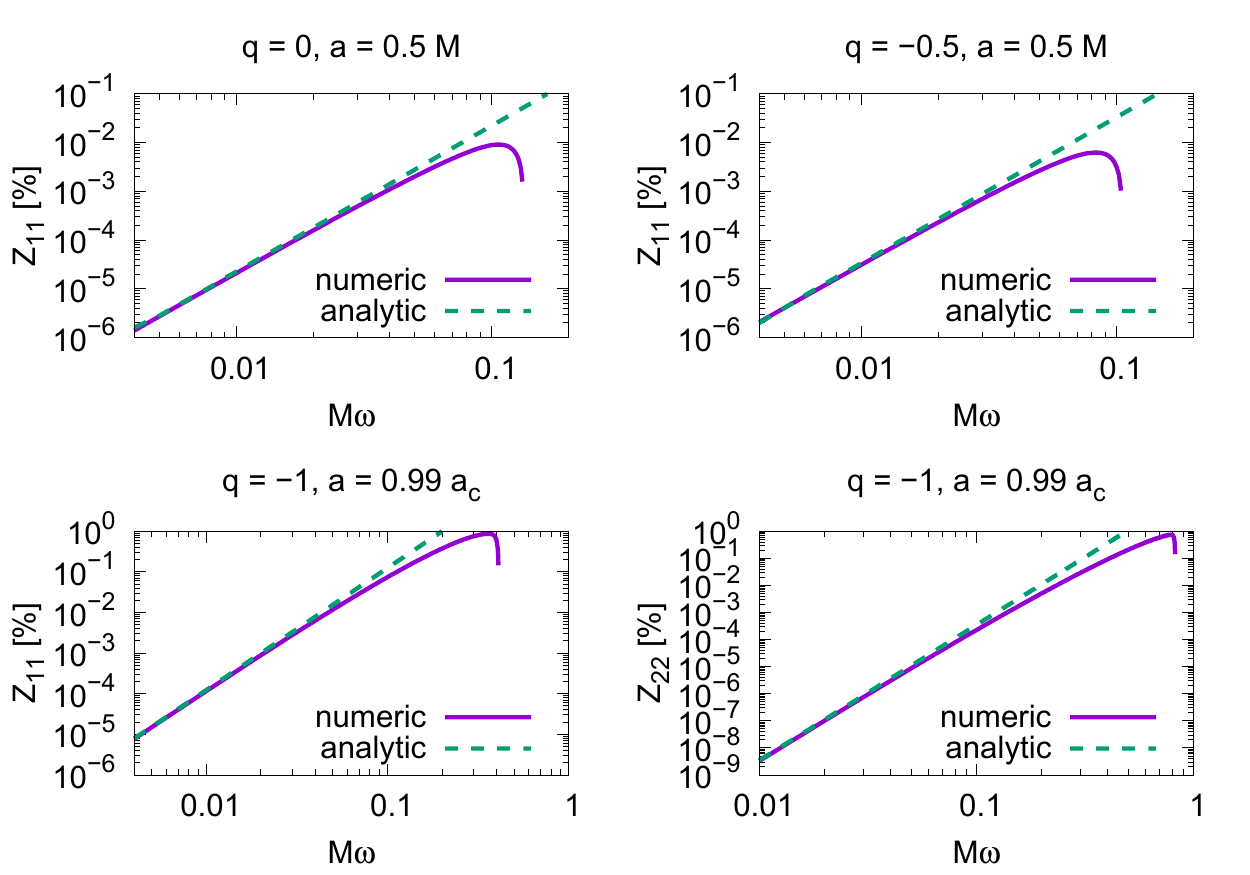}
\caption{A sample of comparisons between the numeric and analytic results obtained in the low-frequency regime. Similar agreement has been found for other choices of the set ($q,a$).}
\label{fig:lf_comp}
\end{figure}

We start by confronting our numeric results against the low-frequency approximation~\eqref{lf_R} in Fig.~\ref{fig:lf_comp} for the cases $q = 0, -0.5$ with $a = 0.5 M$ and $l = m = 1$, and for $q = -1$ with $a = 0.99 a_c$ and $l = m = 1,2$. We see that agreement is excellent regarding $\omega M \lesssim 0.01$. Such good agreement can be considered as a consistency check and precision test of our numeric results. Although graphs in Fig.~\ref{fig:lf_comp} correspond to a very small part of the parameter space in which the low-frequency approximation is valid, we have made similar comparisons for other randomly chosen sets ($q,a$) obtaining similar agreements to the ones presented here.

\begin{figure*}[!htb]
 \centering
 \includegraphics[width=0.98\textwidth]{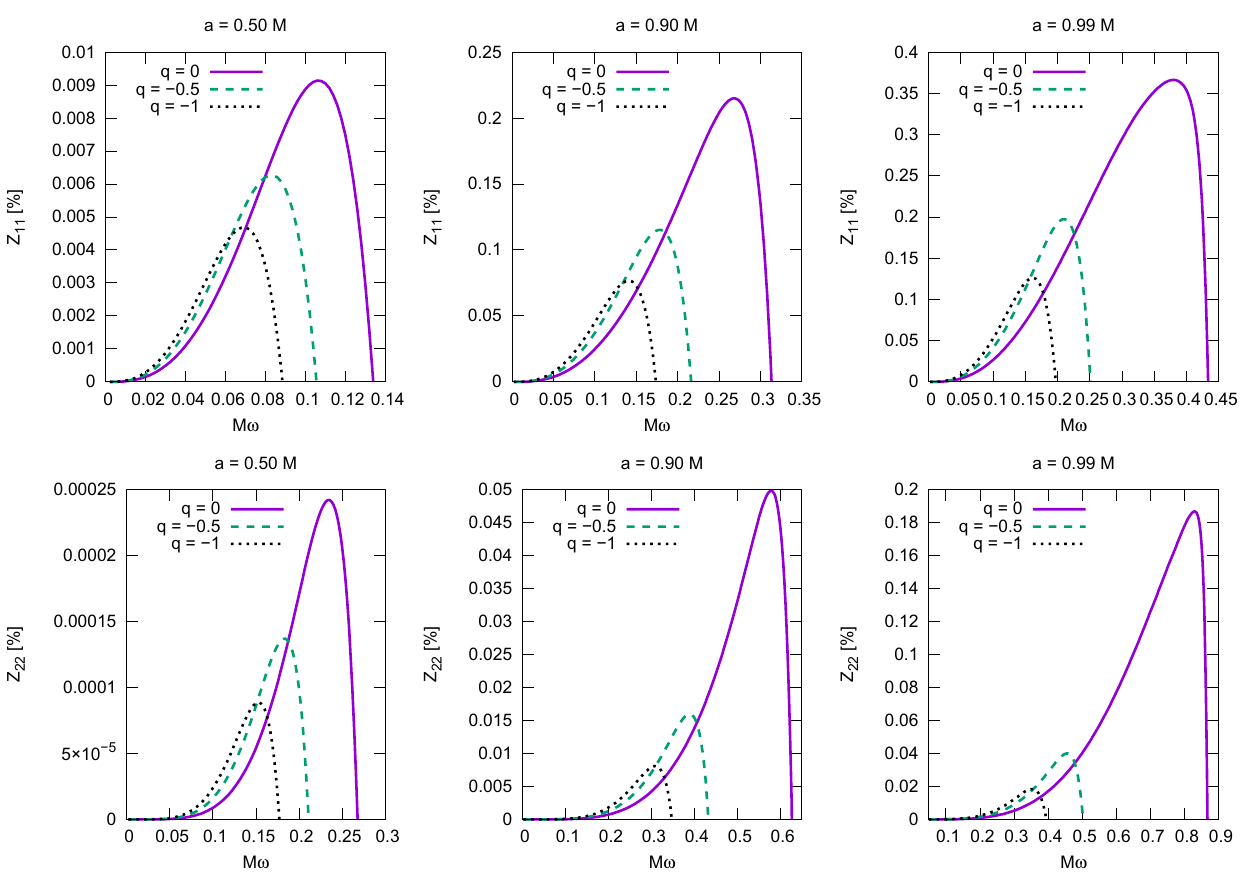}
 \caption{Amplification factors for black holes with $q = 0, -0.5, -1$ for $a = 0.5 M$ (left), $a = 0.90 M$ (center) and $a = 0.99 M$ (right). The considered modes are $l = m = 1$ (top) and $l = m = 2$ (bottom).}
 \label{fig:ref_a0.99}
\end{figure*}

Figure~\ref{fig:ref_a0.99} shows the amplification factor of massless scalar waves scattered by Kerr black holes and by rotating black holes with tidal charges $q = -0.5,-1$ for $a/M = 0.50,0.90, 0.99$, and $l = m = 1,2$. We see that the efficiency of energy extraction from black holes with more intense tidal charges gets suppressed if the ratio $a/M$ is fixed. This is in accordance with the fact that the space between the event horizon and the ergosurface shrinks with the increase of $-q$ for black holes of same $a/M$. We emphasize the fact that in this figure $a$ is in units of the black hole mass, and that $a_c/M = 1, 1.225,1.414$ for $q = 0,-0.5,-1$ respectively. Therefore, when $a/M$ is constant, $a/a_c$ decreases with the increase of $-q$.

\begin{figure}[!htb]
 \centering
 \includegraphics[width=0.48\textwidth]{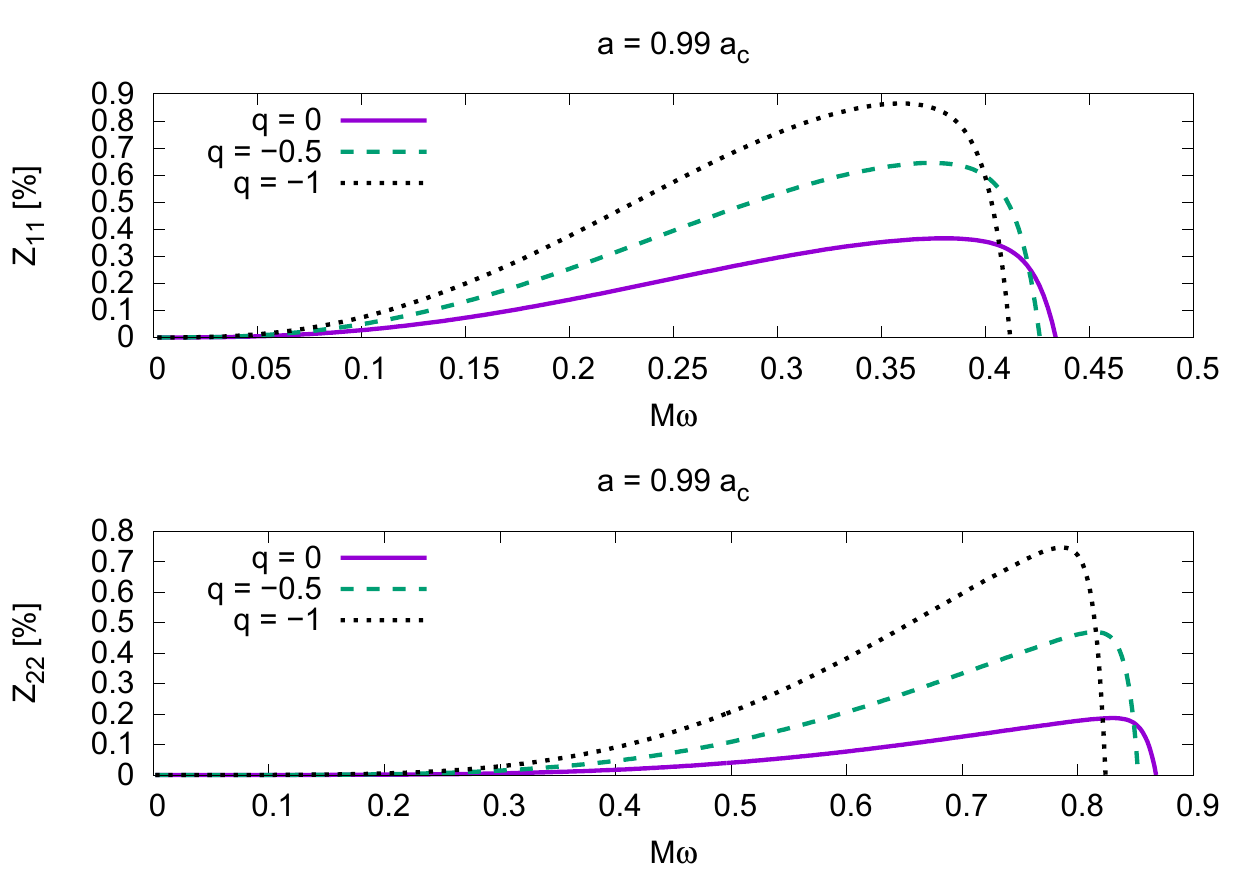}
 \caption{Amplification factors for $a = 0.99 a_c$ in the cases $ q = 0,-0.5,-1$ for $l = m = 1$ (top) and $l = m = 2$ (bottom). In terms of the black hole mass $a/a_c = 0.99$ means $a/M \approx 1.21, 1.40$ for $q = -0.5,-1$ respectively.}
 \label{fig:ref_0.99ac}
\end{figure}

In Fig.~\ref{fig:ref_0.99ac} we show the amplification factors for near-extreme black holes, $a = 0.99 a_c$, with $q = 0, -0.5, -1$ considering modes $l = m =1, 2$. Oppositely to what happens in the case of fixed $a/M$ (Fig.~\ref{fig:ref_a0.99}), now the increase of $-q$ causes maximum amplification to be considerably enhanced. This is a consequence of the fact that extreme black holes with tidal charge can spin faster than Kerr black holes so that $a_c/M$ increases considerably with the increase of $-q$. Moreover, when $a$ is allowed to increase, the event horizon shrinks leaving more space between it and the ergosurface, as illustrated in Fig.~\ref{fig:erg_rh}. We also note that, after reaching a maximum, $Z_{lm} \to 0$ for smaller frequencies once $-q$ increases indicating that superradiance efficiency increases despite the fact $M\Omega_H$ decreases.

\begin{figure}[!htb]
 \centering
 \includegraphics[width=0.49\textwidth]{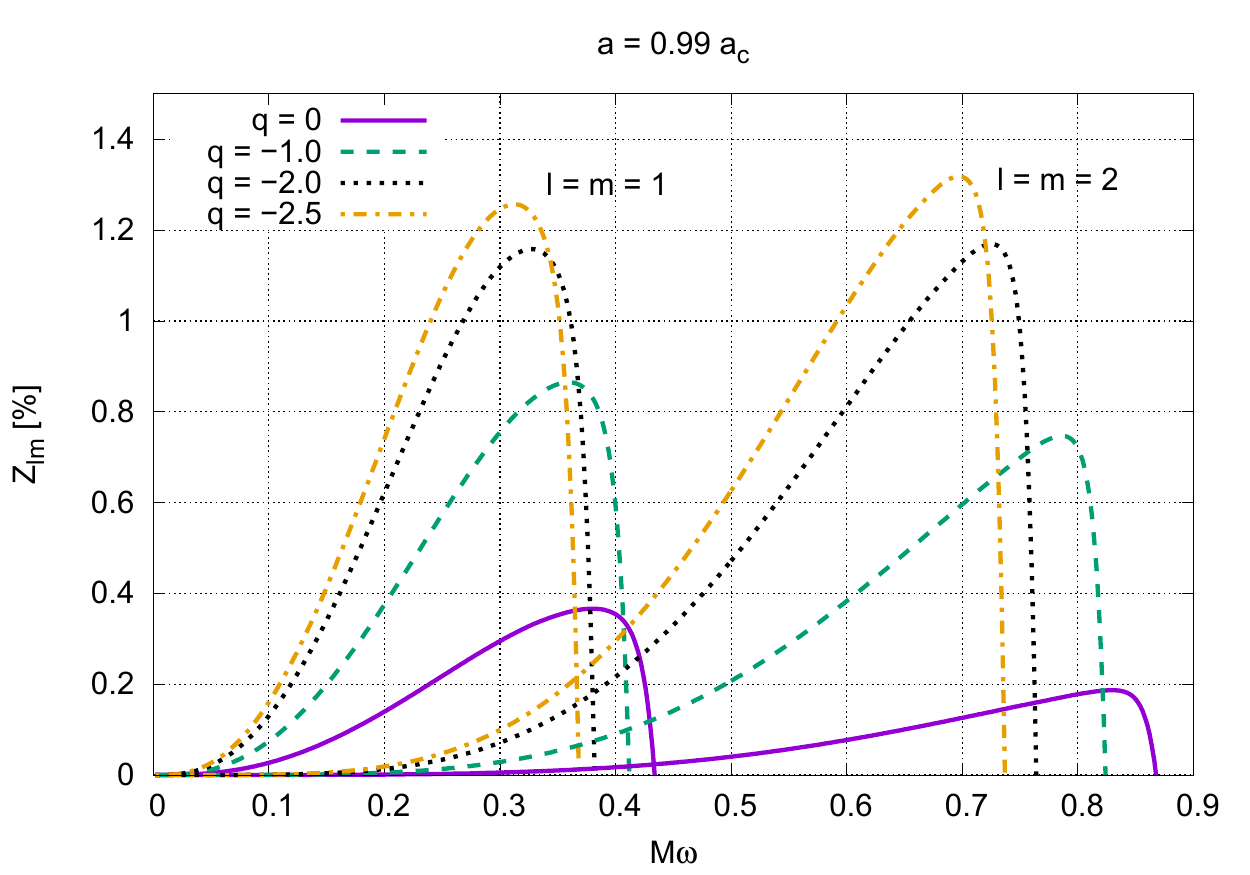}
 \caption{Amplification factors for near-extreme black holes with $q = 0, -1.0,-2.0,-2.5$ and modes $l = m = 1,2$. Grid lines help us to visualize that maximum amplification in the cases $q = -2.0,-2.5$ occurs for $l = m = 2$ instead of $l = m = 1$.}
 \label{fig:ls_comp}
\end{figure}

\begin{figure}[!htb]
 \centering
 \includegraphics[width=0.49\textwidth]{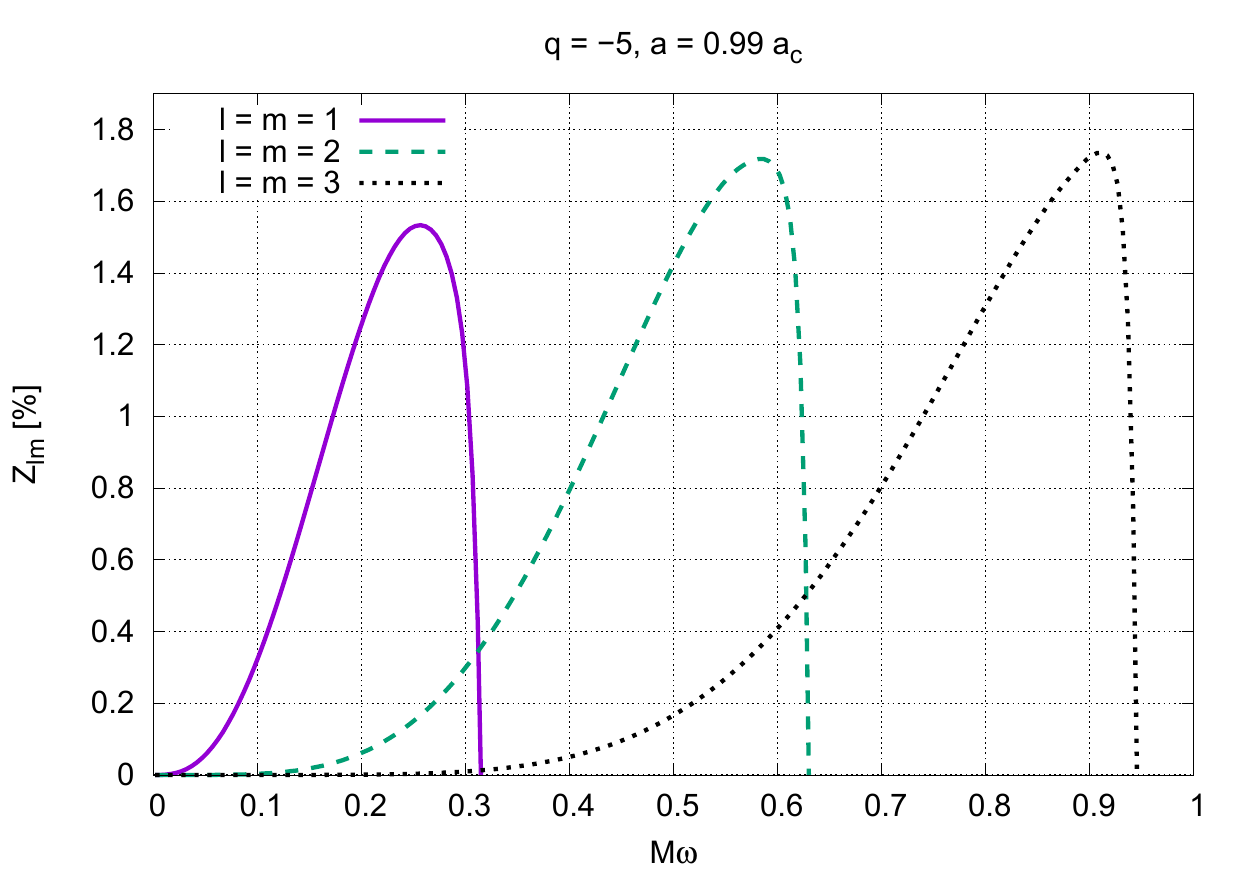}
 \caption{Amplification factors for $q = -5$, $a = 0.99a_c$ and modes $l = m = 1,2,3$. Grid lines help us to verify that in such case maximum amplification occurs for $l = m =3$.}
 \label{fig:amp-qm5}
\end{figure}

The analysis of Fig.~\ref{fig:ref_0.99ac} suggests maximum amplification of modes $l = m = 2$ increases faster than the one of modes $l = m = 1$ as the value of $-q$ increases. In Fig.~\ref{fig:ls_comp} we show superradiance evolution for $l = m = 1$ and $l = m = 2$ considering black holes with $q = 0, -1.0,-2.0,-2.5$. As the value of $-q$ increases, superradiance for $l = m = 2$ enhances considerably so that maximum amplification in this case exceeds the one of modes $l = m = 1$ for $q$ around $-2$. Once the values $q$ can assume have no limit in the negative branch, theoretically, we wonder if maximum amplification could be due to modes $l = m > 2$ for higher values of $-q$. Figure~\ref{fig:amp-qm5} proves this indeed happens indicating that the value of $m$ for which maximum amplification occurs increases with $-q$. There we present amplification factors for $l = m = 1,2,3$ considering near-extreme black holes with $q = -5$.

\subsection{Non-superradiant modes}

\begin{figure}[!tpb]
\centering
 \includegraphics[width=0.49\textwidth]{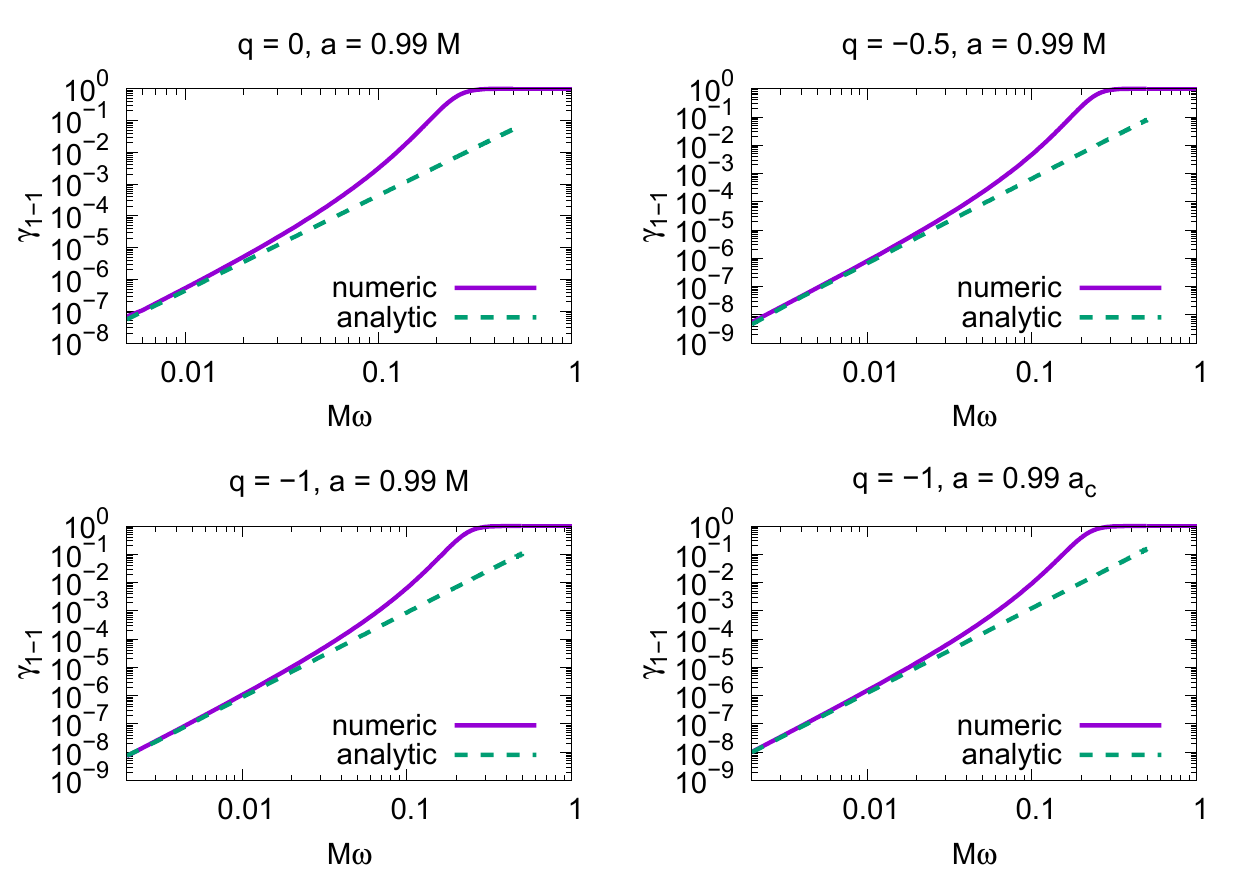}
 \caption{Sample of comparisons between greybody factors obtained analytically in the low-frequency regime and numerically. }
 \label{fig:lf_comp-non_sup}
\end{figure}

Although our analysis focuses on the superradiant regime, it can be straightforwardly applied to compute the greybody factors of non-superradiant modes. Such computation may help to elucidate the physics behind rotating black holes with tidal charge and show that the validity of approximation~\eqref{lf_R} extends to the non-superradiant regime. This approximation is compared with numeric results in Fig.~\ref{fig:lf_comp-non_sup} for a random selection of the parameters $q,a$ with $l = -m = 1$. Similarly to the case of superradiant scattering, the agreement is excellent in the regime of validity of Eq.~\eqref{lf_R}. We have also confronted this analytic expression against other numeric results with similar agreement to the ones illustrated in Fig.~\ref{fig:lf_comp-non_sup}.

\begin{figure*}
 \centering
 \includegraphics[width=\textwidth]{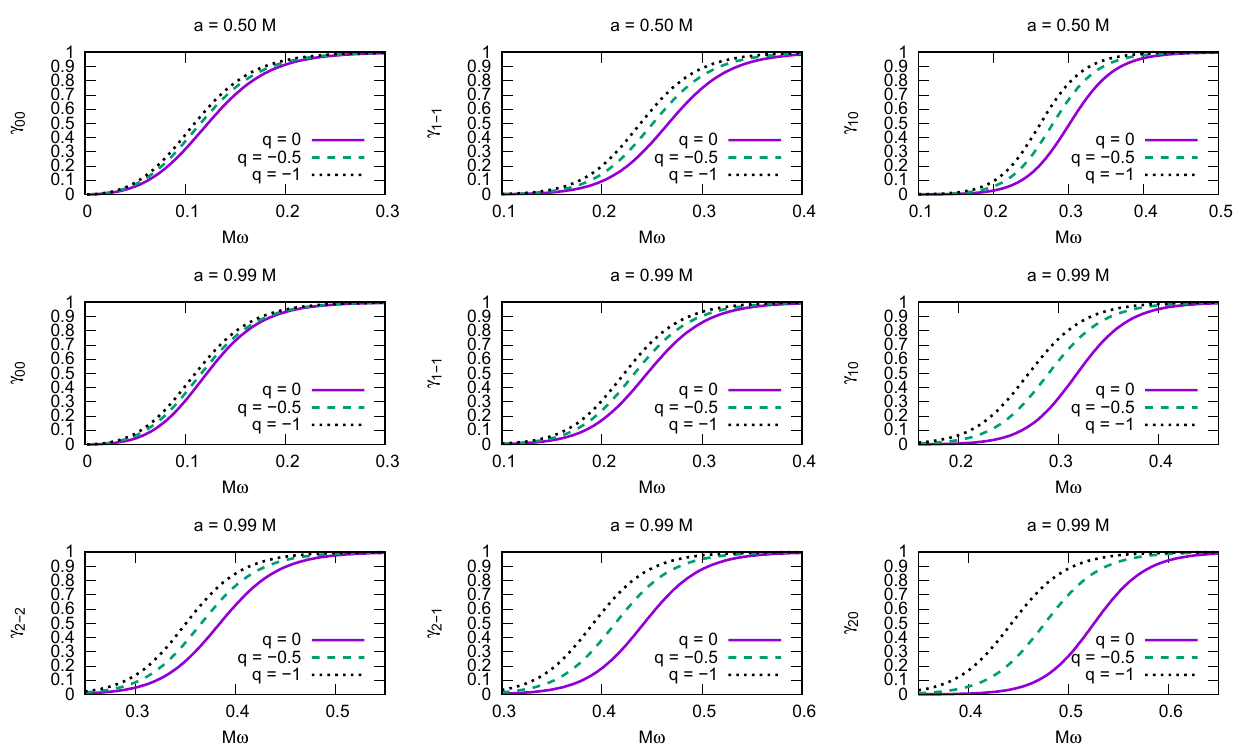}
 \caption{Greybody factors describing the scattering of massless scalar waves by rotating black holes with tidal charge computed numerically for $a/M = 0.50,0.99$, $q = 0,-0.5,-1$, and $l = 0,1,2$.}
 \label{fig:gbf}
\end{figure*}

In Fig.~\ref{fig:gbf} we show the greybody factors for the first non-superradiant modes, $l = 0,1,2$, for $a/M = 0.5, 0.99$ and $q = 0, -0.5, -1$. We see that the effect of increasing $-q$ on the scattering of such modes is to shift their absorption to lower frequencies. This can be understood via the classical analysis of absorption, where the key-feature is the critical impact parameter, $D_c$. Particles in geodesics which have impact parameter $D<D_c$ are absorbed ($\gamma_{lm} \to 1$), while the ones in geodesics with $D > D_c$ are scattered back to infinity ($\gamma_{lm} \to 0$). In the high-frequency limit, the impact parameter is related to $l$ and $\omega$ through $D \sim (l+1/2)/\omega$~\cite{Futterman_etal1988}. It has been shown that the increase of $-q$ results in a increase of $D_c$ considering static cases~\cite{Toshmatov2016prd93_124017,deOliveira2018epjc78_876}. Therefore, the increase of $-q$ results in a transition between scattered to absorbed modes at lower values of $\omega$ considering fixed values of $l$. This analysis not only supports the observed in Fig.~\ref{fig:gbf}, but also sheds light on the reason why maximum amplification occurs for smaller frequencies in Figs.~\ref{fig:ls_comp} and~\ref{fig:amp-qm5} once $-q$ increases.

\section{Final remarks}
\label{sec:conclusions}

We have inferred the influence of a (negative) tidal charge on the phenomenon of superradiance which occurs around rotating black holes by solving the problem of massless scalar waves being scattered by them. Considering black holes with fixed angular momentum per unit mass, the increase of the tidal-charge intensity, $-q$, results in a decrease of amplification of the scattered wave (Fig.~\ref{fig:ref_a0.99}). This is related to the fact that a tidal charge acts to increase the event horizon of the black hole leaving less space between it and the ergosurface.

On the other hand, an increase of the tidal charge intensity results in an enhancement of amplification obtained via the scattering of massless scalar waves around near-extreme rotating black holes (Fig.~\ref{fig:ref_0.99ac}). This accomplishment is only possible, however, because more negative tidal charges allow black holes to spin faster than extreme Kerr black holes~\cite{Aliev2005prd71_104027}.

We have also shown that superradiance efficiency around near-extreme black holes increases despite the fact that the angular velocity at the event horizon decreases with the increase of $-q$. This is only possible because maximum amplification is shifted to smaller frequencies as $-q$ increases (Figs~\ref{fig:ref_a0.99}, \ref{fig:ref_0.99ac}, and~\ref{fig:ls_comp}), similarly to what happens with greybody factors for non-superradiant modes (Fig.~\ref{fig:gbf}). One way of understanding such behavior can be done by making use of the semiclassical relation $D \sim (l+1/2)/\omega$. Once ergoregion increases with $-q$, waves with higher impact parameter can access it, causing superradiance to occur for smaller frequencies if we consider fixed values of $l$.

It is well known that maximum amplification is achieved for $l = m = 1$ in the scattering of massless scalar waves by Kerr black holes~\cite{Brito2015lnp906}. Considering this, the tidal charge imposes a crucial change at superradiance, once it may lead maximum amplification to be obtained from modes $l = m = 2, 3$ (Figs.~\ref{fig:ls_comp} and~\ref{fig:amp-qm5}). Although our analysis has been done for $q \ge -5$, it indicates that the increase of $-q$ can lead maximum amplification to be obtained even for higher values of $m$.

We provided an analytical description for the reflection coefficients valid at low frequencies [Eq.~\eqref{lf_R}]. This approximation fits our numeric results for both superradiant and non-superradiant modes with $m \neq 0$ once $M\omega \lesssim 0.01$ (Figs.~\ref{fig:lf_comp} and ~\ref{fig:lf_comp-non_sup}). This agreement shows our results are consistent and it allowed us to use the analytical result to calibrate the precision in our numeric computations.

\section*{Acknowledgments}

The author would like to thank Conselho Nacional de Desenvolvimento Científico e Tecnológico (CNPq) for partial financial support via the grants 304679/2018-6 and 427532/2018-3.

\end{document}